\documentclass[11pt,a4paper,dvips]{article}

\usepackage{amstex}
\usepackage{latexsym}
\usepackage{graphicx}
\usepackage{amssymb}

\def\apj{ApJ}
\def\apjs{ApJS}
\def\mnras{MNRAS}

\def\opt{$d \tau / d z \; $}
\def\ltk{\left [ \,}

\def\rtk{\, \right  ] }

\def\kms{km~s~$^{-1}$ }

\def\N#1{{N({\rm #1})}}
\def\cm#1{\, {\rm cm^{#1}}}

\def\lya{Ly$\alpha$ }
\def\sigfc{\sigma_{f_c}}
\def\delv{\Delta v}
\def\sci#1{{\rm \; \times \; 10^{#1}}}
\def\Nperp{N_\perp}

\begin{document}
\baselineskip24pt
\pagestyle{plain}

\begin{center}
\vspace{0.2in}

{\Large \bf Damped Lyman Alpha Systems at High Redshift and}\\
\bigskip 
{\Large \bf Models of Protogalactic Disks}\\
\vspace{.3in}

Karsten Jedamzik$^1$ {\it and} Jason X. Prochaska$^2$

$^1$Max-Planck-Institut f\"ur Astrophysik

85740 Garching bei M\"unchen
\medskip

$^2$
Department of Physics, and Center for Astrophysics and Space Sciences,

University of California, San Diego 

\end{center}

\vspace{0.1in}

\medskip

We employ observationally determined intrinsic velocity widths
and column densities of
damped Lyman-alpha systems at high redshift to investigate 
the distribution of baryons in protogalaxies within the context
of a standard cold dark matter model.
We proceed under the assumption that damped Lyman alpha systems 
represent
a population of cold, rotationally supported, protogalactic
disks and that the abundance of dark matter halos is well
approximated by a cold dark matter model with critical density
and vanishing cosmological constant. 
Using conditional cross sections
to observe a damped system with a given velocity width and
column density,
we compare observationally inferred velocity width and column density
distributions to the corresponding theoretically determined
distributions for a variety of disk parameters and CDM normalizations.
In general, we find that the observations can not be reproduced by the
models for most disk parameters and CDM normalizations. 
Whereas the column density
distribution favors small disks with large neutral gas fraction, 
the velocity width
distribution favors large and thick disks 
with small neutral gas fraction.
The possible resolutions of this problem in the context of this CDM model
may be: (1) an increased 
contribution of rapidly rotating disks within massive dark matter halos to 
damped Lyman-alpha absorption or (2) the abandoning of simple
disk models within this CDM model
for damped Lyman-alpha systems at high redshift.
Here the first possibility may be achieved by supposing that
damped Lyman-alpha system formation only occurs in halos with
fairly large circular velocities and the second possibility may result from
a large contribution of mergers and double-disks to damped
Lyman-alpha absorption at high redshift.

\medskip

\newpage
\pagestyle{plain}

\section{Introduction}

It is widely believed that the phenomena of damped Lyman-alpha
(hereafter; damped Ly-$\alpha$)
absorption
in the spectra of high redshift quasars is due to large amounts of neutral
gas in dark matter halos which intervene on the lines of sight to
quasars \cite{wol86}. 
It is intriguing that the ratio of neutral gas mass density
to the critical density
in damped Ly-$\alpha$ systems, 
$\Omega_{Ly\alpha }(z)$, at redshift $z\sim 3$ is comparable
to the baryonic content of stars in galaxies at the present epoch
\cite{lzwt95,wol95}. 
Both the steady decline in $\Omega_{Ly\alpha }(z)$ with decreasing redshift
and the agreement between $\Omega_{Ly\alpha }(z)$ at 
low $z$ and the neutral gas content of nearby galaxies \cite{rao93}
are commonly interpreted as the incorporation of baryons into stars 
within the damped systems \cite{lzwt95,wol95}. 
Observations of damped Ly-$\alpha$
systems have, and continue, to yield important information about the 
abundances and masses of dark matter halos, as well as the process
of star formation \cite{ma94,kau94,mom94}.

Whereas the hosts of damped Ly-$\alpha$ systems seem
to be established, there is
considerable debate about the distribution of neutral gas
in individual dark matter halos at high redshift. For example, 
neutral gas in damped Ly-$\alpha$ systems may be distributed 
spherically symmetric within the halo, or may
be within a rotationally supported protogalactic disk.
The main observational support for damped Ly-$\alpha$
systems representing halo gas comes from a comparison
of metallicities of damped Ly-$\alpha$ systems as a function
of redshift with the metallicities of disk and halo stars as a
function of stellar age \cite{ptt94,lu96b}. 
On the other hand, there is accumulating
observational evidence that 
damped  Ly-$\alpha$ systems
at high redshift are protogalactic disks.
Prochaska \& Wolfe (1997a; hereafter PW) 
recently analyzed a sample of 17 damped Ly-$\alpha$ systems
observed at high-resolution with HIRES on the Keck I telescope. 
They argue that
observationally determined velocity profiles of high-redshift
damped systems are reflective of rapidly rotating, cold
disks, and cannot be reconciled with isothermal halo 
gas models or slowly rotating \lq\lq hot\rq\rq disks.
In addition, the recent observation of the stellar and continuum
emission of a high-redshift galaxy believed to be responsible for
damped Ly-$\alpha$ absorption in the spectrum of a quasar
may be explained straightforwardly only if one assumes the galaxy
to be host of a disk \cite{djg96,lu97}.

Unfortunately it is not possible to easily obtain a resolution
to the debate about the spatial and velocity distribution of neutral gas 
in dark matter halos from numerical simulations. This is partially due to 
limited resolution, such that simulations which contain enough
halos to be statistically representative may not resolve
the physics within an individual halo \cite{grd97}. Furthermore, when
simulating the formation of disks in individual halos within the
context of currently popular structure formation scenarios, the sizes
of disks are predicted to be much smaller than the observationally
inferred sizes of present day galaxies, casting significant doubt on
the predictive power of the simulations \cite{nav95}. 
Nevertheless, numerical simulations may shed further light on the 
nature of damped  Ly-$\alpha$ systems; for example, they may test
if it is realistic to employ simple models for the
velocity distribution of gas or if, rather, the occurrence of damped systems 
results from cold, rotating disks, as well as halo gas, merging galaxies, 
and/or irregular galaxies in almost equal parts \cite{hae97}.

Semianalytic models for the formation of disk galaxies are
very successful in explaining properties of present-day spiral galaxies
\cite{fal80,kau96,dal97,mo97}.
These models consist of two parts, a Press-Schechter type
formalism to determine the abundances and masses of halos
in a given hierarchical structure formation scenario, and a model for
the formation of a rotationally supported disk within the dark halo.
The sizes and profiles of disks may be estimated by the distribution of
halo angular momentum under the assumption that there is no
angular momentum transport between halo and disk, and that
baryons are initially in solid body rotation. When combined with
efficiencies to turn gas into stars, such models successfully explain
the observed trends and scatter in exponential disk scale length, surface
brightness, and the Tully-Fisher relationship \cite{dal97,mo97}. 
It is instructive to see if such disk models may also 
reproduce the observational data of damped Ly-$\alpha$ systems at high
redshift. In fact, an initial, not fully self-consistent, 
comparison between the observational data
and one particular disk formation model \cite{kau96} demonstrated 
that there is
significant discrepancy between the model
and the observations (PW).

In this paper we compare observationally inferred velocity widths
of a sample
of damped Ly-$\alpha$ systems at mean redshift $z\approx 3$ to theoretically
predicted velocity widths for a variety of disk models. 
In Section 2 we summarize the assumptions entering our calculation
of halo abundances
and masses, as well as disk rotational
velocities. We also describe our computation
of conditional cross sections, to observe 
a disk with more than the critical column density for a damped system
and a velocity width more than a certain fraction of the rotational velocity of
the disk. Our results and conclusions are presented in Section 3.

\section{Semianalytic Model of Protogalactic Disk Populations}

We consider the conditional optical
depth per unit redshift, $d\tau / dz (\Delta v_{th})$,
for observing
a damped Ly-$\alpha$ system 
showing a velocity spread, $\Delta v$, larger than some
threshold, $\Delta v_{th}$. Note that the optical depth per unit redshift
is frequently called the rate of incidence.
The conditional optical depth
is the product of proper number density of halos hosting damped systems,
the conditional cross section for observing a damped system
with velocity spread larger than $\Delta v_{th}$, and
the proper path length traveled by a photon per unit redshift interval,
i.e. $d\tau /dz = n^p\sigma_d dl^p/dz$. We assume that the kinematics
and gas distribution of damped Ly-$\alpha$ 
systems is adequately described by rotationally supported disks 
within dark matter halos.
To account for a continuous distribution
of halo masses we use the Press-Schechter formalism 
for halo formation with density threshold
$\delta_c=1.68$ and top hat filter. 
In our calculations we assume that the abundances
of halos are well approximated by the abundances in a closed cold dark matter
model with vanishing cosmological constant and varying normalizations.
We use the CDM transfer function by Bardeen {\it et al.} (1986) with
scale invariant initial perturbations $(n=1)$ and Hubble parameter
$H_0=50$ km s$^{-1}$
Mpc$^{-1}$ ($h=0.5$). 
The conditional optical depth for a continuous mass distribution
may be written as

\begin{equation}
{d\tau\over dz}({\Delta v_{th}})={c\over H_0}(1+z)^{-5/2}
{\int_{M(v_c^{min})}^{\infty}dM {dn^p(M)\over dM}
\sigma_d(M,\Delta v_{th})}\, ,\label{eq:1}
\end{equation}
with $c$ the speed of light.
Note that we will use an equation similar to Eq.~\ref{eq:1} for
the computation of column density distributions where
the conditional disk cross section in Eq.~\ref{eq:1} is replaced
by an adequate column density distribution.
In writing Eq.~\ref{eq:1} we assume that damped systems only
form within sufficiently large halos with circular velocities exceeding
a minimum value, $v_c^{min}$. Even though it has been argued on
physical grounds that $v_c^{min}\approx 35-50$ km s$^{-1}$ due
to photoionization by the UV-background at high redshift and the
associated long cooling time scales for baryons in small halos 
\cite{efs92},
we will treat $v_c^{min}$ as a free parameter in order to assess the
contribution of less massive halos to the conditional optical depth.

We may parametrize the conditional disk cross section by

\begin{equation}
\sigma_d(M,\Delta v_{th})=\pi R_d^2 \, F(N_{\perp}, \kappa \, |
\, N\geq N_d,
f_v=\Delta v/v_{rot}\geq \Delta v_{th}/v_{rot})\, ,\label{eq:2}
\end{equation}
where $R_d$,  $N_{\perp}$, $\kappa$, and $v_{rot}$ are scale length, 
central column density, ratio of
scale height to scale length, 
and rotational velocity of the disk, respectively. 
The cross section receives only contributions
from lines of sight which (1) result in column densities $N$ exceeding
the threshold for a damped system $N_d=2 \sci{20} \cm{-2}$ and
(2) show velocity width $\Delta v=f_vv_{rot}$ exceeding the velocity width
threshold, $\Delta v_{th}$. Note that due to the self-similarity of disks with
different $R_d$ but the same $N_{\perp}$ and $\kappa$,
the function $F$ depends explicitly only on $N_{\perp}$ and $\kappa$,
but not $R_d$, provided one assumes a flat rotation curve.
We assume the disk
rotational velocity equals the circular velocity of the halo 

\begin{equation}
v_{rot}=v_c=\sqrt{G M\over r_{vir}}=159.6 {\rm km\over s}h^{1/3}\biggl
({M\over 10^{12}M_{\odot}}\biggr) \bigl(1+z\bigr)^{1/2}\,  .\label{eq:3}
\end{equation}
Here $r_{vir}$ is the virial radius of a halo calculated within the
spherical collapse model, and $G$ is the gravitational constant.
Motivated by the successes of semianalytic formation models of 
present-day disk
galaxies, we assume that the scale length
of the disk is proportional to the virial radius of the halo

\begin{equation}
R_d=f r_{vir}={f\over (179)^{1/3}}\biggl({3M\over 4\pi\rho_0}\biggr)^{1/3}{1\over
(1+z)}=169.7{\rm kpc} \;
f {h^{-2/3}\over (1+z)}\biggl({M\over 10^{12}M_{\odot}}
\biggr)^{1/3}\, .\label{eq:4}
\end{equation}
In order to evaluate conditional disk cross sections, we make several 
simplifying assumptions regarding the physical
characteristics of the disks.  We adopt a simple, 
phenomenologically motivated disk model, the exponential disk.
As in PW, we parameterize the neutral gas density of the disk, $n_H$,
by

\begin{equation}
n_H(R,Z) = n_0 \exp \ltk - {R \over R_d} - {|Z| \over \kappa R_d}\rtk\, ,
\label{TRDvol}
\end{equation}
where $n_0$ is the central gas density.
Furthermore, we describe the gas kinematics 
by a flat rotation curve, parameterized by the
rotation speed, $v_{rot}$, and a random motion component
characterized by a 1-dimensional velocity dispersion, 
$\sigma = 10$ \kms. 
Conserving baryonic mass within halos, the central column density
may be related to the total halo mass by

\begin{equation}
N_\perp = \int n_H(0,Z) \; dZ = 2 n_0 \kappa R_d =
{f_N \Omega_b M\over \mu m_H 2\pi R_d^2}\, ,  \label{Ncolm}
\end{equation}
where $m_H$ is the hydrogen mass, $\Omega_b=0.05$ is the 
assumed fractional
contribution of baryons to the critical density, and $\mu\approx 1.3$
accounts for the fact that a fraction of all baryons are in form of helium.
The parameter $f_N\leq 1$ is the fraction of baryons participating
as neutral gas in the disk. This fraction may be smaller than one
if either there is an ionized component of gas due to partial photoionization
or if gas is efficiently incorporated into stars.

In order to calculate the cross section $\sigma_d(M,\Delta v_{th})$
we consider two conditional
cross-sections: (1) $\sigma_{NHI}$, defined as
the cross-section for sightlines yielding column densities
$\N{HI} \geq 2 \sci{20} \cm{-2}$ (the damped \lya threshold), and
(2) the conditional velocity width
cross-section, $\sigfc$, for sightlines which give a velocity
width, $\delv$, greater then a fraction, $f_c$, of the
rotation speed ($f_c \; \epsilon \; [0,1]$).
We calculate $\sigfc$ and $\sigma_{NHI}$
by systematically running sightlines through an exponential
disk according to the techniques outlined in $\S 3.2$ of PW.
While computer intensive, the process is straightforward
and reproducible.  
To verify the accuracy of our procedures we compare $\sigma_{NHI}$
with the cross-sections determined by Fall $\&$ Pei (1993) in the
thin-disk limit and find excellent agreement
over the whole column density range $\Nperp = 10^{18.9} 
- 10^{23.9}\cm{-2}$.
Note also that we find the thin-disk limit to be
an excellent approximation
to $\sigma_{NHI}$ even for finite thickness disks ($\kappa\sim 0.5$).
To simplify the computations, we
make use of the fact that $\sigfc$ is independent
of $\Nperp$. 
We may then obtain the doubly conditional cross section $\sigma_d$ 
for given $\Nperp$, $\kappa$ and $f_c$ by convolving
the cross-sectional area leading to $\sigma_{NHI}$ with the
cross-sectional area leading to $\sigfc$.  

The doubly conditional cross-section depends on the disk parameters in the 
following ways.
Thicker disks (larger $\kappa$ values) imply
larger velocity widths and therefore larger $\sigma_d$
for a given $f_c$ value due to the increased path length
of sightlines within the disk. The dependence
of $\sigma_d$ on $\Nperp$ is more complicated.  
As noted in PW (e.g. Figure 3),
the velocity widths tend to decrease with larger impact parameters
for a given sightline. 
This is primarily because the differential
rotation projected along the line of sight decreases with
increasing impact parameter.  For a disk with a large $\Nperp$ value
even sightlines 
at a  number of scale lengths may lead to the observation of a damped system.
However, due to the large impact parameter such sightlines only yield 
fairly small $\Delta v/v_{rot}$.
These ideas are illustrated in Figure 1 where we
plot $\sigma_{d}$ for two different $\Nperp= 10^{20.5}, 10^{22.1}$cm$^{-2}$ and
$\kappa = 0.108, 0.232$ as
a function of $f_c$, and where $\sigma_{d}
= \sigma_{NHI}$ for $f_c \leq 0.05$.
It is essential for the conclusions of this paper that we find only a small
fraction $\sim 1/10$ of sightlines which result into the observation of a 
damped system also show more than $\sim$50\% 
of the rotation velocity
for large $\Nperp$ disks. This is not the case for small $\Nperp$ disks
where this ratio is approximately one-half.

\section{Results}

Given the results from $\S2$, we can calculate the optical depth
per unit redshift, \opt, for damped \lya systems showing more
than a certain velocity width, $\delv_{th}$, as a function of $\delv_{th}$.
The model has 5 free parameters: (1) $\sigma_8$, the linear
rms fluctuation amplitude measured at 8 h$^{-1}$ Mpc, 
(2) $v_c^{min}$, the circular velocity cutoff corresponding to the least massive
dark matter halo capable of forming a damped system,
(3) $\kappa$, the ratio of the scale height to the scale length, 
(4) $f$, the scale length of the disks as a fraction of the virial radius,
and
(5) $f_N$, the fraction of baryons within
a dark matter halo which contribute to the disk.
In Figures 2a-e we compare optical depth curves
calculated over a wide range of the physically allowed 
parameter space with \opt measured from 26 damped \lya
systems with redshifts $z = 2.0 - 4.4$ ($\bar z = 2.86$).  Fourteen
of the systems were presented in PW and the remaining 12 can be found 
in Prochaska $\&$ Wolfe (1997b).  The sample is kinematically
unbiased and each system has an accurately determined
velocity width.

Figures 2a and 2b present the effects of $\sigma_8$ and $v_c^{min}$, respectively,
on the predicted optical depth.  These two parameters determine the
number density of halos as a function of circular velocity.  
In particular, by increasing $\sigma_8$ and/or $v_c^{min}$ the mass
function of halos capable of forming damped systems is shifted towards
halos with larger $v_c$, and disks with larger $v_{rot}$. 
In panel (a) we consider a range of $\sigma_8$ values,
including the two most favored observational values: 
$\sigma_8 = 0.667$ from the observed masses of rich clusters
\cite{whte93} and $\sigma_8 = 1.2$,  the 
normalization implied by measurements of the CMBR \cite{gor96}.
First note that none of the \opt curves in panel (a) are consistent
with the observational data.
Whereas the observations indicate that a fraction of one-half of all
damped systems show an intrinsic velocity width of $\delv\geq 100$ km s$^{-1}$
our models show that this fraction is $\leq 1/10$ for $\sigma_8\leq 1$.  
Note that we choose $f=0.1$ and $f_N=1$ in panel (a) yielding typical central
column densities of $\Nperp\sim 10^{22}$cm$^{-2}$. The small fraction
of large $\delv$ systems in our model is then due 
almost in equal parts to a comparatively large fraction
of small $v_c$ systems as well as due to the small chance for observing
a significant fraction of the rotational velocity at such high $\Nperp$ (cf. Fig. 1).
It is evident from panel (b) that by removing the contributions
of small halos ($v_c\leq 200$\kms) to the optical depth
one may obtain approximate agreement between observations and
the model.
This highlights the fact that the CDM model with $\sigma_8\approx 1$
predicts enough massive halos ($v_c\geq 200-300$km s$^{-1}$) 
at high redshift to account for the abundance
of damped systems with large intrinsic velocity width.

The dependence of the optical depth on the
parameters $\kappa$, $f$ and $f_N$ which affect the cross-section 
of the disks within the dark matter halos, is shown
in panels (c)$-$(e).  In Figure 2c we plot
\opt against $\delv$ for models with varying disk thickness.
It is seen that the observational data favors thick disks and that
disks with thicknesses comparable to the thin disk of the 
Milky Way ($\kappa\approx 0.05$; Mihalas \& Binney 1981) are unacceptable.  
We investigate the effect of varying the disk size in panel (d).  
One may produce a nearly acceptable fit of the velocity
distribution data even for low normalization
CDM models ($\sigma_8=0.5$) by taking the scale length of the
disks to be a considerable fraction of the virial radius ($f\approx 0.8$).
Agreement between the model predictions for the velocity width
distribution and the observations is then attained due to the
comparatively large typical fraction of the rotational velocity one
observes in these low central column density ($\Nperp\sim 10^{20.3}$cm$^{-2}$)
systems (cf. Fig. 1).
It is important to stress that analytic models of disks within a
CDM cosmogony (e.g. Dalcanton et al.\ 1997, Mo et al.\ 1997)
suggest $f \approx \bar\lambda$, the average spin angular momentum parameter
which has been measured numerically to
be $\bar\lambda \approx 0.05 \pm 0.03$ \cite{barnes87}.
Such analytic models are thus inconsistent with extremely large scale length $R_d$.
Finally, we allow for the possibility that not all of the baryons
within the halos will collapse to form a gaseous disk.  
Numerical simulations suggest \cite{ma97} 
that a fraction of baryons in halos
will remain hot due to photoionization
by the UV-background and therefore not contribute to damped \lya 
absorption. Moreover, it is possible that a significant fraction of baryons
may be incorporated into stars by a redshift of $z\approx 2.5$.
As is evident from panel (e) in Fig. 2, lowering $f_N$ has the effect of
increasing the fraction of high $\delv$ systems, thereby slightly lessening
the discrepancy between model and observations. This is again due to
a decrease in the average central column densities of disks.  

Nevertheless, disks with low central column densities are inconsistent with
the column density distribution for
damped \lya systems at high redshift. In Fig. 3 we compare the column
density distribution resulting from several of the models in Fig. 2
to the  observationally determined 
column density distribution at mean redshift $\bar z = 3.0$
\cite{wol95}. It is evident from the figure that the observational
data clearly favors large $\Nperp$ disks 
(the dotted and long-dashed lines are models with low average $\Nperp$).
Significantly increased scale lengths $R_d$,
or decreased neutral fractions $f_N$, are therefore ruled out as an
explanation for the comparatively flat velocity width distribution
of damped  \lya systems. 

While it seems difficult to simultaneously reproduce the velocity width
and column density data of damped \lya systems
within the context of simple disk models in a
CDM cosmogony, our model has enough parametric freedom to yield
an acceptable fit. In Figure 2 (panel f) and Figure 3 
(the dashed-dotted line) 
we show such a model with fairly thick disks ($\kappa =0.39$), 
large velocity cutoff ($v_c^{min}=150$km s$^{-1}$), 
and small neutral fraction ($f_N=0.33$). 
However, such a fit requires significant ``fine tuning'' and yields 
a physically unmotivated model.
As such, we believe that simple
disk models in a CDM cosmogony cannot reproduce straightforwardly
the observational data of damped \lya systems at high redshift. This is
in contrast to the success of such models in explaining the
properties of present day spiral galaxies (e.g. Kauffmann 1996). 
Nevertheless, it is
interesting to note that CDM models for the formation of structure,
combined with assumptions of gas cooling and star formation, tend
to significantly overproduce the abundances of faint galaxies 
\cite{wht91}. 
Though there are alternative explanations to this apparent discrepancy
between observations of the present-day luminosity function 
and the model predictions,
one solution would be a suppression of gas cooling and star formation in
low circular velocity halos. If cooling of gas in less massive 
$v_c\leq 200$km s$^{-1}$ halos
would be similarly suppressed at high redshift, than protogalactic disks within
a CDM cosmology would remain as a viable model for the nature of damped
\lya systems at high redshift. 

There are two physical effects which have not been taken into account in
our model and which could alleviate the discrepancy between the
observationally determined velocity width distribution and our model 
predictions.
Mo et al.\ (1997) 
compute rotation curves for galactic disks within realistic dark
matter halos. They find that low $v_c$ halos have rotation curves which
have a peak of approximate height $v_{rot}\approx 1.5v_c$.
Since this peak is approximately located at a
distance from the disk center where damped \lya absorption is dominated 
(at the threshold for damped systems $N_d$) the \lq\lq effective\rq\rq 
rotation velocity of low $v_c$ halos may exceed $v_c$. Further, hydrodynamic
simulations employed in the analysis of Ma et al.\ (1997)
indicate that there may
be partial ionization of disks at larger distances from the disk center. 
Lines of sight which penetrate the disk at large impact 
parameters typically yield only small
$\delv$. If those lines of sight do not contribute to damped \lya absorption,
the average observed $\delv$ of a given disk increases. It remains to be
shown if these effects can significantly improve the comparison between
the observationally determined velocity width distribution and the disk model
predictions.

In summary, we have compared observationally determined
velocity width and column density distributions of damped \lya systems
at high redshift to theoretically predicted distributions within the context 
of protogalactic disk models for damped \lya systems in a closed CDM model
with vanishing cosmological constant. Whereas the column density distribution
may be reasonably well reproduced by assuming rotationally supported
disks with scale length comparable to those inferred from successful 
semianalytic models for the formation of present-day spiral galaxies,
the simplest protogalactic disk models cannot reproduce the velocity
width distribution. This discrepancy is mainly due to the large fraction
of small circular velocity halos within the CDM model. Approximate
agreement between the observational data and the disk model predictions
may be obtained if the contribution of low circular
velocity systems $v_c\leq 200$ km s$^{-1}$ to damped \lya absorption
is significantly diminished. By increasing the thickness of disks and slightly
decreasing the neutral fraction of gas one may further improve the agreement
between observational data and the model. Leaving the realm of the
model investigated in this paper, there are two alternative solutions to the
problem. It may well be that only a fraction of all damped systems
are \lq\lq regular\rq\rq protogalactic disks. In this case most of the damped
\lya systems which show large intrinsic velocity width
$\delv\geq 100$ km s$^{-1}$ have to be the result of damped \lya
absorption resulting from lines of 
sight through double disks, ongoing mergers, 
and irregular gas distributions. Since the observed fraction of damped
systems with $\delv\geq 100$ km s$^{-1}$ is large ($\sim 50\%$) only a similarly
large fraction of such systems at high redshift $z\approx 3$ may reproduce
the observations.
Further light on this issue may be shed by hydrodynamical simulations
of structure formation [e.g. Haehnelt et al.\ 1997]. 
Finally, it may simply be that the CDM models
employed in this paper
overproduce the abundances of small $v_c$ halos and do not represent
the \lq\lq correct\rq\rq structure formation scenario.

\vskip 0.2in

We acknowledge many useful discussions with Martin Haehnelt,
Houjun Mo, Simon White, and Arthur Wolfe. JXP would especially like
to thank W.L.W. Sargent and L. Lu for providing their HIRES
spectra.  This work was supported
by NASA grant NAGW-2119 and NSF grant AST 86-9420443 (JXP) and...

\break

\section{Figure Captions}

\centerline{\bf Figure 1}
The doubly conditional cross-section, $\sigma_d$ 
in units of $R_d^2 / \pi$ for a disk yielding a velocity
width $\delv$ greater then a fraction, $f_c$, of the rotation
speed for disks with two different central column densities
($\Nperp = 10^{20.5}, 10^{22.1}$;
top and bottom lines respectively) and 
thickness ($\kappa = 0.108, 0.232$; dashed and solid lines
respectively).

\centerline{\bf Figure 2}
Optical depth per unit redshift, $d\tau /dz$, for damped Ly-$\alpha$ systems
showing a velocity width, $\Delta v$, more than a threshold 
$\delv\geq\Delta v_{th}$, as a function of
$\Delta v_{th}$. The observationally determined distribution
is shown by $\pm 1\sigma$ bars for a sample of 26 damped Ly-$\alpha$ systems
with mean redshift $z=2.87$. The dotted lines show theoretically determined 
velocity width distributions at the same redshift for cold, rotating disks
in CDM models with Hubble constant 
$h=0.5$ and
baryonic contribution to the closure density, $\Omega_b=0.05$. 
Panel (a) corresponds to models with parameters $f=0.108$, $\kappa =0.232$,
$f_N=1$, $v_c^{min}=50$km s$^{-1}$, 
and varying CDM normalization $\sigma_8=2, 1, 0.6667, 0.5$ (from
top to bottom, respectively), where $\sigma_8$ is the linear rms fluctuation
amplitude on a scale of 8$h^{-1}$Mpc, $f$ is the exponential disk scale length
in units of the halo virial radius, $\kappa$ is the ratio of scale height to
scale length, $f_N$ is the fraction of baryons contributing to neutral disk
gas, and $v_c^{min}$ is the lowest circular velocity of halos where gas may still
cool and form damped Ly-$\alpha$ systems. 
The remaining panels correspond to models with parameters
(b) $f=0.108$, $\kappa =0.232$, $\sigma_8=1$, $f_N=1$, and
varying velocity cutoff $v_c^{min}=50, 100, 200, 300$ km s$^{-1}$
(from top to bottom, respectively);
(c) $f=0.108$, $\sigma_8=1$, $f_N=1$, $v_c^{min}=50$ km s$^{-1}$, and
varying disk thickness $\kappa =0.5, 0.232, 0.108, 0.05$ (from top to bottom,
respectively); (d) $\kappa =0.232$, $\sigma_8=0.5$, 
$f_N=1$, $v_c^{min}=50$ km s$^{-1}$, and varying
disk size $f=0.834, 0.5, 0.232, 0.108$ (from top to bottom at high $\Delta v$,
respectively);
(e) $f=0.108$, $\kappa =0.232$, $\sigma_8=1$, $v_c^{min}=50$ km s$^{-1}$,
and varying neutral gas fraction $f_N=1, 0.5, 0.25, 0.1$ (from top to bottom,
respectively); and (f) $f=0.108$, $\kappa =0.39$, $\sigma_8=1$, 
$v_c^{min}=150$ km s$^{-1}$, and $f_N=0.33$.

\centerline{\bf Figure 3}

Column density distribution for damped Ly-$\alpha$ systems, $f(N)$, as
a function of $N$ at mean redshift $z=3$. The crosses show observational
data taken from Wolfe et al (1995) with the vertical lines indicating
the $\pm 1\sigma$ ranges. 
The different lines show theoretically determined 
column density distributions for some of the models in Figure 2.
The lines correspond to model parameters $f=0.05$, $\kappa =0.232$,
$\sigma_8=1$, $f_N=1$, and $v_c^{min}=50$ km s$^{-1}$ (solid line); 
$f=0.834$, $\kappa =0.232$,
$\sigma_8=2$, $f_N=1$, and $v_c^{min}=50$ km s$^{-1}$ (dotted line);
$f=0.108$, $\kappa =0.232$,
$\sigma_8=1$, $f_N=1$, and $v_c^{min}=200$ km s$^{-1}$ (short-dashed line);
$f=0.108$, $\kappa =0.232$,
$\sigma_8=1$, $f_N=0.1$, and $v_c^{min}=50$ km s$^{-1}$ (long-dashed line);
and
$f=0.108$, $\kappa =0.39$,
$\sigma_8=1$, $f_N=0.33$, and $v_c^{min}=150$ km s$^{-1}$ (dashed-dotted line);

\break

\begin{figure}
\centering
\includegraphics[height=7.0in, width=5.0in]
{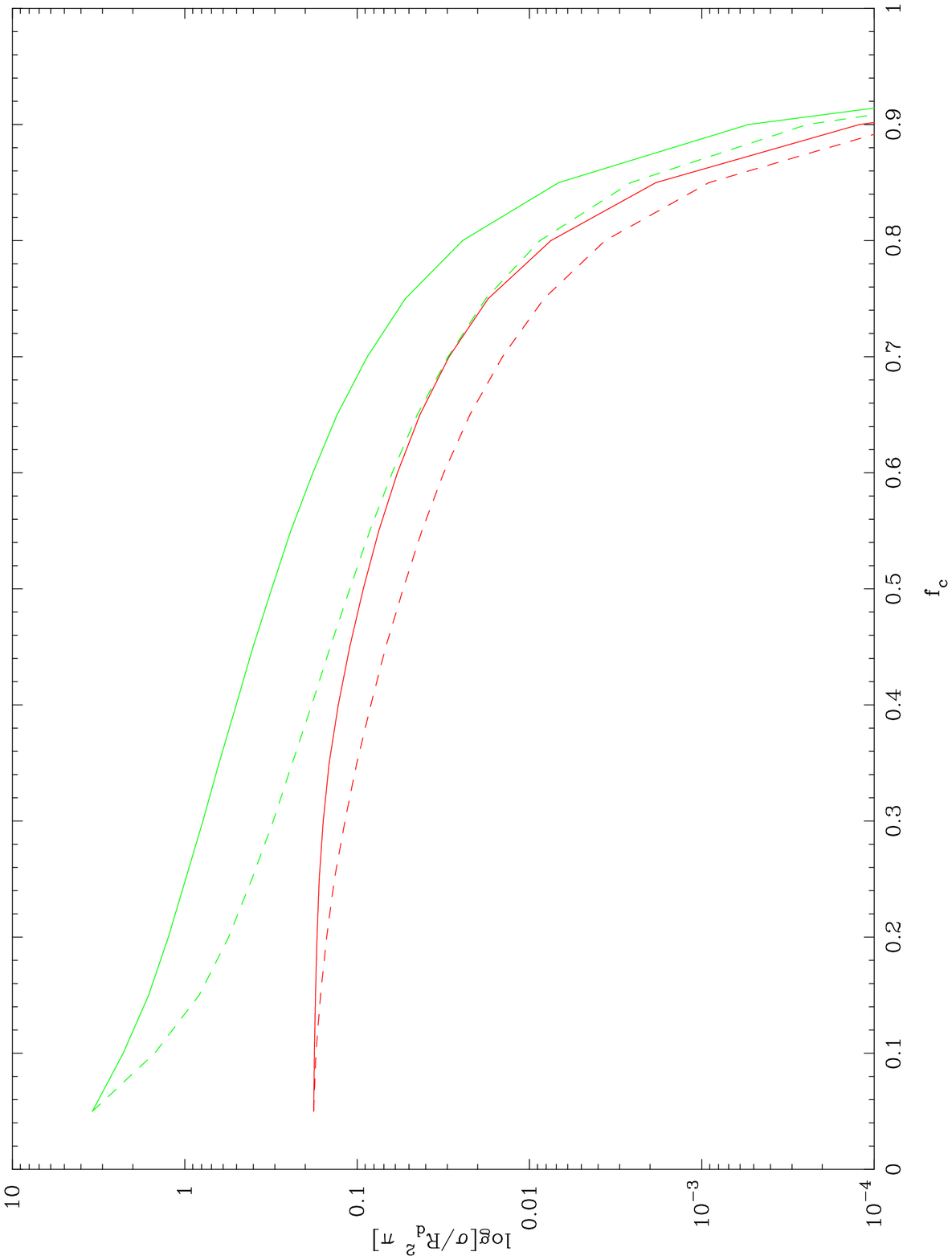}
\caption{}
\end{figure}

\break

\begin{figure}
\centering
\includegraphics[height=7.0in, width=5.0in]
{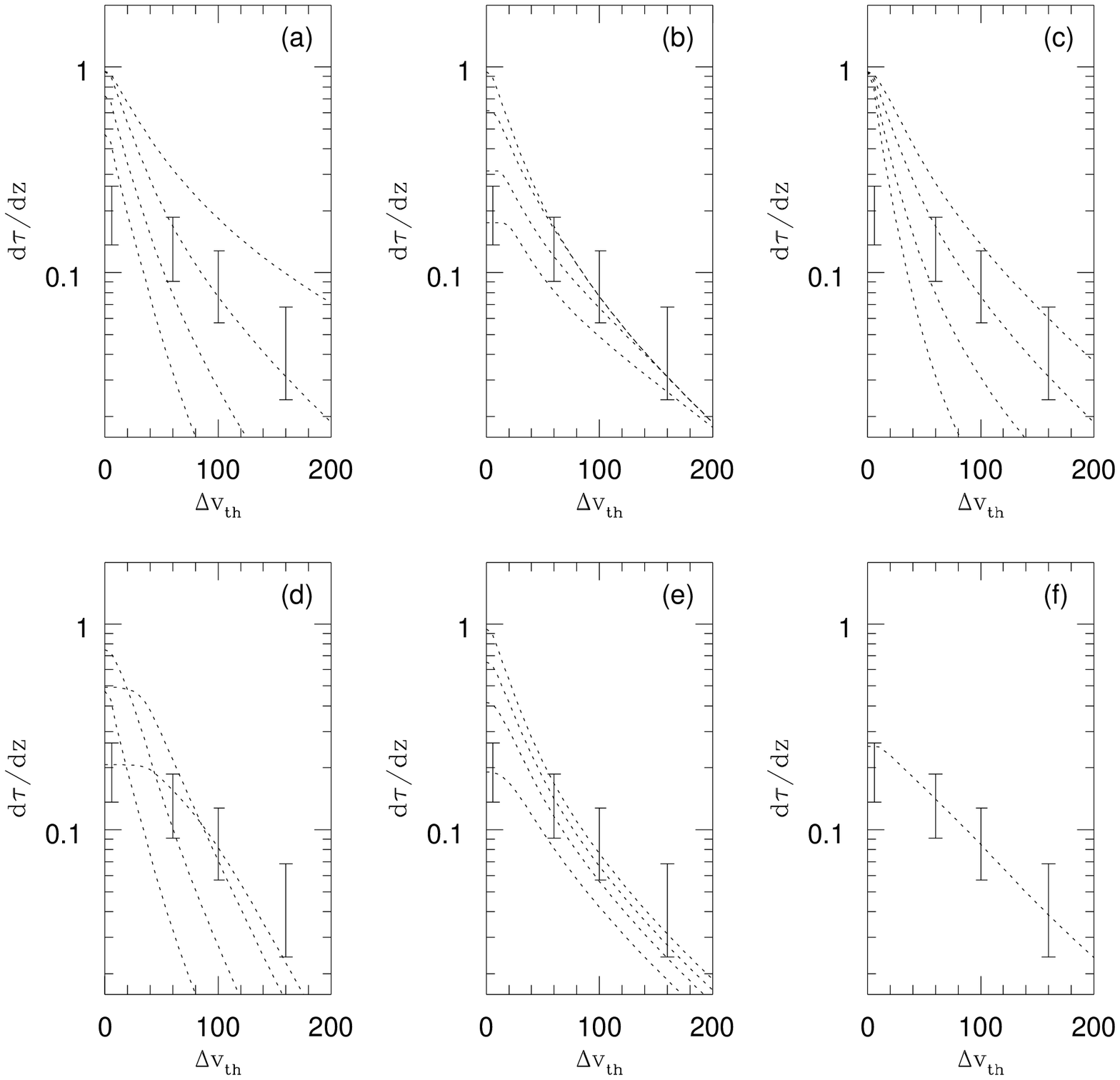}
\caption{}
\end{figure}

\break

\begin{figure}
\centering
\includegraphics[height=7.0in, width=5.0in]
{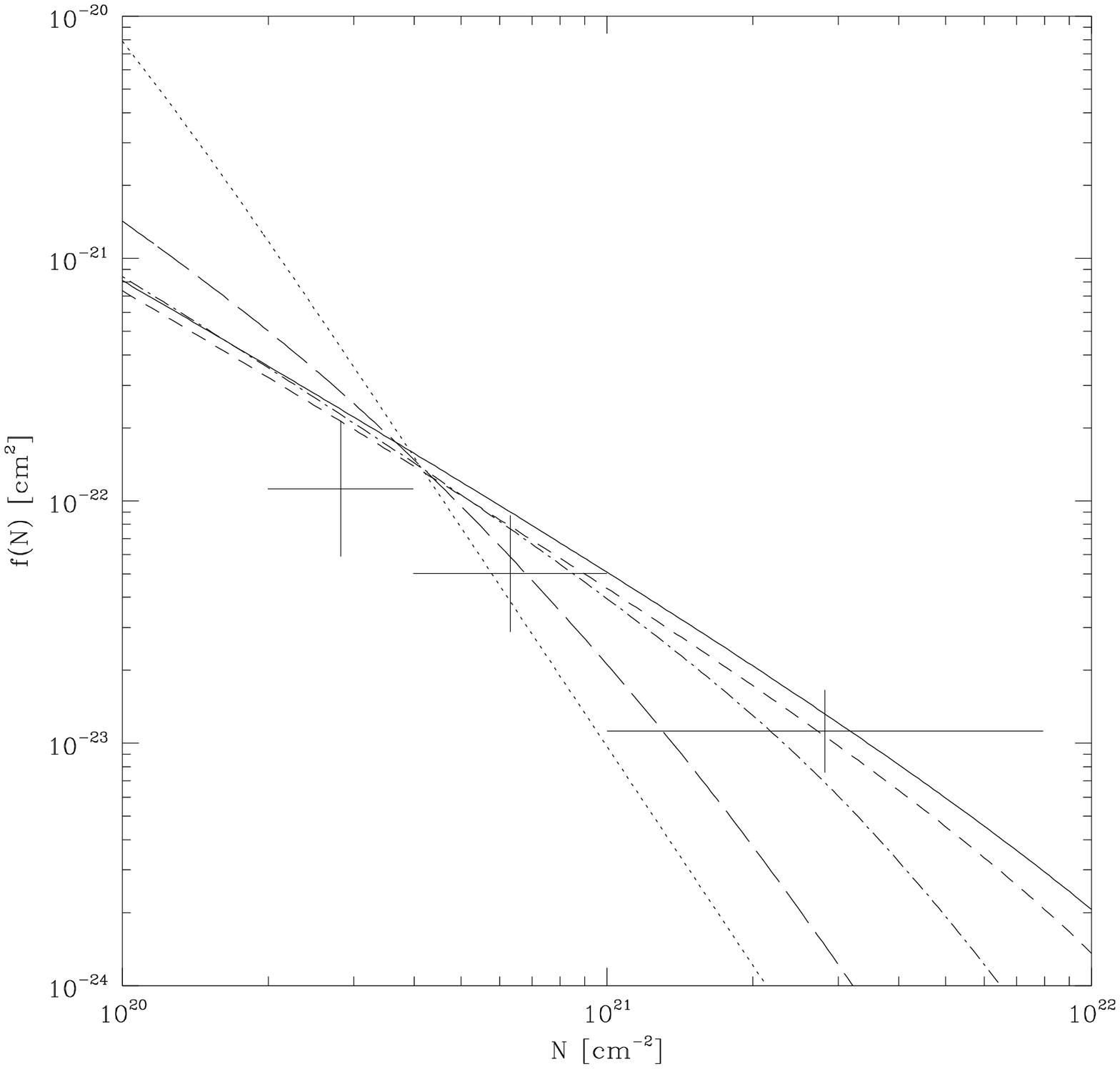}
\caption{}
\end{figure}

\end{document}